\documentclass[final,5p,times,twocolumn]{elsarticle}

\usepackage{amsmath,amssymb,mathtools}
\usepackage{mathrsfs}
\usepackage{bm}
\usepackage{graphicx}
\usepackage{booktabs}
\usepackage{siunitx}
\usepackage{microtype}
\usepackage{hyperref}
\usepackage{xcolor}
\usepackage{enumitem}

\ifdefined\CNSNSSubmission
  \usepackage{lineno}
\fi

\journal{Communications in Nonlinear Science and Numerical Simulation}

\begin{document}

\begin{frontmatter}

\title{Stability and Interaction Dynamics of Solitons in Spatially Engineered High-Order Nonlinear Media}

\author[inst1]{Dao Quang Anh}
\author[inst1]{Bui Duc Tinh} 
\author[inst2]{Doan Quang Tri} 
\author[inst3]{Nguyen Thi Dung} 
\author[inst4]{Duong Chinh Cuong} 
\author[inst5]{Marek Trippenbach} 
\author[inst6]{Nguyen Luong Thien}
\author[inst2,*]{Nguyen Viet Hung }
\affiliation[inst1]{
    organization={Faculty of Physics, Hanoi National University of Education},
    city={Hanoi},
    addressline = {136 Xuan Thuy Str.},
    country={Vietnam}
}
\affiliation[inst2]{organization={School of Materials Science and Engineering (SMSE),Hanoi University of Science and Technology (HUST)},            addressline = {No 1 - Dai Co Viet Str.}, 
city={Hanoi},
country={Vietnam}}
\affiliation[inst3]{organization={Faculty of Natural Sciences, Hong Duc University},            addressline = {565 - Quang Trung Str.}, 
city={Thanh Hoa},
country={Vietnam}}
\affiliation[inst4]{organization={Faculty of Electrical and Electronic Engineering, Phenikaa University},  addressline = {Nguyen Trac Str.}, 
city={Hanoi},
country={Vietnam}}
\affiliation[inst5]{organization={Faculty of Physics, University of Warsaw},            addressline = {Pasteura 5}, 
city={Warsaw, 02-093},
country={Poland}}
\affiliation[inst6]{organization={Viet Nam National Space Center - Viet Nam Academy of Science and Technology}, addressline = {18 Hoang Quoc Viet - Nghia Do}, 
city={Hanoi},
country={Vietnam}}
\affiliation[*]{organization={Email: hung.nguyenviet1@hust.edu.vn}}

\begin{abstract}
We study spatial solitons and their interaction dynamics in nonlinear optical media with spatially engineered refractive index and competing cubic--quintic (CQ) nonlinear profiles, using the variational approximation (VA), the hybrid variational approximation (HVA), and direct numerical simulations. The model was implemented in a symmetric step-index planar dielectric waveguide with a core exhibiting competing CQ nonlinearity and cladding layers possessing only a cubic nonlinear response. For stationary states, the Gaussian VA predicts two types of $N(\mu)$ characteristic curves, where $N$ is soliton norm and $\mu$ is propagation constant, separated by a boundary surface in parameter space, and numerical calculations reveal the same two types. Below this surface, the VA agrees with the numerical results mainly at low powers, while pronounced deviations in the profile and stability appear at high powers. Above the surface, the variational and numerical curves retain the same qualitative form, and a super-Gaussian ansatz accurately describes the high-power flat-top solitons. Soliton collisions produce four post-interaction regimes: Oscillation, Molecular, Splitting, and Breakup. The HVA reproduces the first three over a broad power range, including collisions involving flat-top solitons. Its main limitation arises in the Breakup regime, where strong radiation leaves the guiding region and cannot be represented by the adopted HVA ansatz. Nevertheless, for solitons associated with the second type of characteristic curves, the HVA still captures the breakup dynamics qualitatively. Thus, the HVA provides an efficient description of complex soliton interactions at a substantially lower computational cost than direct numerical simulations.
\end{abstract}


\end{frontmatter}

\ifdefined\CNSNSSubmission
  \linenumbers
\fi

\section{Introduction}
Spatial optical solitons are self-localized light beams sustained by a balance between transverse diffraction and the nonlinear refractive-index change induced by the beam itself \cite{kivshar2003optical}. Their particle-like nature enables them to interact via the nonlinear index landscape generated during propagation \cite{science.286.5444.1518,RevModPhys.61.763}. In integrable systems, soliton collisions are fully elastic \cite{RevModPhys.61.763}; in non-integrable settings, including multidimensional models and systems with higher-order nonlinearities, they may involve radiation, energy exchange, trapping, fusion, or soliton-molecule formation \cite{RevModPhys.61.763,PhysRevLett.76.2698,QI20122372,PhysRevE.110.044215,2fqh-mhs7,10.1063/1.5034361,10.1063/5.0309512,PhysRevE.94.032217}.

A standard nonintegrable setting in nonlinear optics is the competing cubic--quintic (CQ) model, combining a self-focusing cubic term with a self-defocusing quintic term \cite{PhysRevLett.88.073902}. In homogeneous media it supports multidimensional fundamental solitons, including three-dimensional spatiotemporal solitons (light bullets) \cite{PhysRevLett.88.073902}. Localized states have also been studied in spatially structured systems, including linearly coupled media \cite{DROR2011526}, dual-core configurations \cite{PhysRevE.67.056608}, double-well potentials \cite{PhysRevA.84.053618,PhysRevE.110.064216,PhysRevA.90.023841}, and periodic lattices \cite{PhysRevE.91.023203}. Beyond fundamental solitons, CQ media support vortex states \cite{WENG201866,PAREDES2022133340,Dong:23}, dipoles and quadrupoles \cite{Zeng2025}, elliptical and rectangular solitons \cite{ZENG2024114645}, flat-top solitons \cite{PhysRevE.110.044215,2fqh-mhs7}, and multi-peak excited states \cite{PhysRevE.68.046612}.

A key feature of the CQ model is the saturation of Kerr self-focusing at high intensity. In a pure Kerr medium, beams may undergo critical collapse in two dimensions and supercritical collapse in three dimensions once the power exceeds the corresponding threshold \cite{Fibich:00,PhysRevLett.90.203902}. The self-defocusing quintic term arrests unlimited narrowing and permits stable localized states \cite{PhysRevLett.88.073902}, although it does not make every CQ solution stable \cite{PAREDES2022133340}.

The same nonintegrability also makes CQ-soliton interactions highly diverse. Collisions may generate radiation and energy exchange, or lead to trapping, fusion, and bound states \cite{PhysRevE.110.044215,2fqh-mhs7,10.1063/1.5034361,10.1063/5.0309512,PhysRevE.94.032217,TATSAGOUM2025108528,PhysRevE.88.062904,Chen:14}. While stationary CQ solitons have been studied extensively, their collision dynamics, especially for flat-top states, remain less explored. The outcomes are also sensitive to the initial conditions and physical parameters, so a systematic map of the dynamical regimes remains of interest.

Fifth-order optical nonlinearities have been measured in several media, notably carbon disulfide (CS$_2$) \cite{Kong_2009,Besse2014CS2}, fused silica \cite{Ekvall2001Silica,Poezzhalov2026Silica}, and optical glasses \cite{Chen2006}. Chalcogenide glasses also exhibit strong Kerr responses, as established by Z-scan and pump--probe measurements \cite{Smektala2000Chalcogenide,Gopinath2004GeAsSe}. In CS$_2$, the competition between self-focusing cubic and self-defocusing quintic nonlinearities has been used to create stable two-dimensional spatial solitons \cite{PhysRevLett.110.013901,PhysRevA.102.033523}.

Because the CQ model is nonintegrable, exact analytical solutions are available only in special cases; most studies therefore rely on approximate analytical techniques or numerical computation. Stationary states can be obtained, for example, by the Newton conjugate-gradient method (NCGM), modified squared-operator method (MSOM), or accelerated imaginary-time evolution method (AITM), while their stability can be tested by linear stability analysis or direct split-step Fourier (SSFM) propagation \cite{yang2010nonlinear,agrawal2019nonlinear}.

The variational approximation (VA) provides a complementary analytical route and has been widely used for optical solitons \cite{PhysRevA.27.3135}, optical switching \cite{MALOMED200271}, laser-beam collapse \cite{Desaix:91}, stationary states and dynamics of light bullets \cite{PhysRevE.70.016603,PhysRevLett.10.1103}, and Bose--Einstein condensates \cite{PhysRevA.59.620,PhysRevE.82.046602,HUNG201437,VIETHUNG20091449}. With a suitable trial function (ansatz), the VA reduces nonlinear partial differential equations to equations for a finite set of collective coordinates, providing approximate information on existence, stability, and dynamics. Its accuracy, however, is controlled by the chosen ansatz and may deteriorate when interacting nonlinear waves deform strongly or emit radiation.

The hybrid variational approximation (HVA), introduced by Edwards and co-workers \cite{Edwards2005Hybrid}, relaxes this restriction. Instead of prescribing an ansatz for the full wave field, it applies a trial profile only along the confined coordinate(s), while retaining unknown functional dependence along the remaining directions. This keeps the variational reduction but allows substantially more freedom during the evolution. The HVA has been used for Bose--Einstein condensates in optical lattices \cite{Edwards_2005_2}, condensate expansion in ring-shaped potentials \cite{PhysRevE.86.056710}, Bragg interferometry with ultracold atoms \cite{PhysRevA.84.043648}, propagation and interaction of spatial solitons in Kerr graded-index waveguides \cite{VI2024108124}, and pulse splitting in inhomogeneous optical media \cite{Infeld_2006,Infeld2006}. Its application to CQ systems, particularly to the formation and interaction of flat-top solitons (FTs), has received much less attention.

Here we consider spatial solitons in a higher-order nonlinear medium where both the refractive index and the CQ coefficients are modulated in one transverse direction. The model corresponds to a symmetric planar waveguide with a nonlinear core, in which self-focusing cubic and self-defocusing quintic responses compete, surrounded by cladding layers with a weaker quintic contribution. A step-index contrast provides transverse confinement. We first determine the existence and stability of stationary solitons and then systematically study two-soliton collisions as the initial conditions are varied. Numerical results are compared with VA and HVA predictions to establish the accuracy and range of validity of the variational description and to construct maps of the post-collision dynamical states.

The paper is organized as follows. Sec.~II introduces the model and its possible experimental realization. Sec.~III analyzes stationary solitons and their stability using the VA and numerical methods. Sec.~IV addresses soliton interactions by comparing the HVA with direct simulations. Sec.~V summarizes the main results.

\section{The model}
\begin{figure}[!tb]
    \centering
    \includegraphics[width=\linewidth]{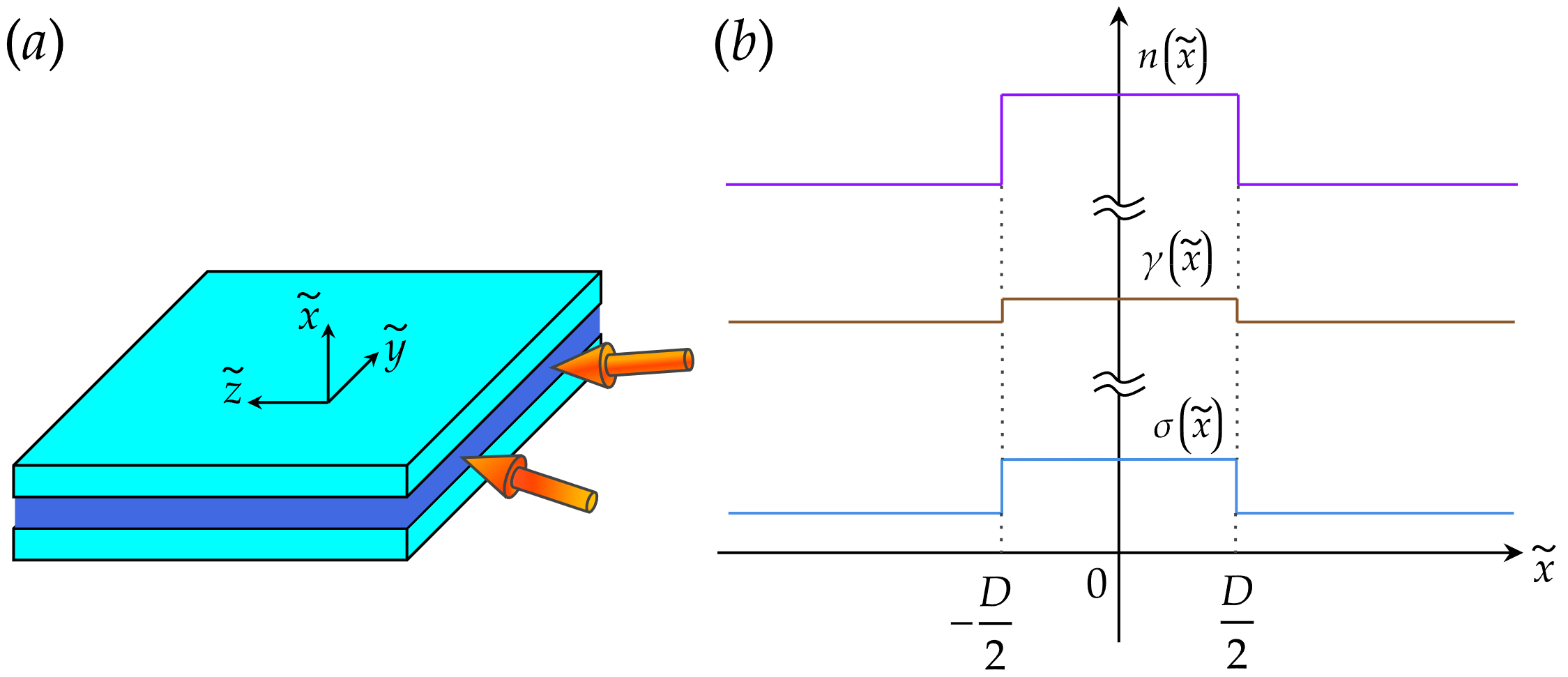} 
    \caption{(a) Schematic of the single-core step-index waveguide, with the high-index core $(n_{\mathrm{c},i})$ sandwiched between lower-index cladding layers $(n_{\mathrm{cl},i})$. The input beam propagates along $z$ and is confined to the core. (b) Step-index profile along the $\tilde{x}$ axis.}
    \label{fig.1}
\end{figure}

We consider optical-beam propagation in a higher-order nonlinear medium whose linear and nonlinear indices are modulated along one transverse direction. A symmetric dielectric planar waveguide provides a direct realization of this setting, as sketched in Fig.~\ref{fig.1}. Within the paraxial approximation, retaining the nonlinear polarization up to fifth order \cite{boyd2020nonlinear}, the slowly varying complex envelope $A(\tilde{\mathbf{r}},\tilde{z})$ obeys
\begin{equation}
    i\frac{\partial A}{\partial\tilde{z}}+\frac{1}{2k_0n_{\mathrm{cl},0}}\tilde{\Delta} A+k_0\left(\delta n_0(\tilde{x})+n_2(\tilde{x})|A|^2+n_4(\tilde{x})|A|^4\right)A=0,
    \label{eq:1}
\end{equation}
where $\tilde{\Delta}=\partial^2/\partial\tilde{x}^2+\partial^2/\partial\tilde{y}^2$, $\tilde{\mathbf{r}}=(\tilde{x},\tilde{y})$ denotes the transverse coordinates, and $\tilde{z}$ is the propagation coordinate in physical units. Here $k_0$ is the vacuum wavenumber, $n_{\mathrm{cl},0}$ is the cladding refractive index, $\delta n_0(\tilde{x})=n_0(\tilde{x})-n_{\mathrm{cl},0}$ is the local linear-index contrast, and $n_2$ and $n_4$ are the cubic (Kerr) and fifth-order nonlinear coefficients, respectively. The linear and nonlinear coefficients are taken in the stepwise form
\begin{align}
n_i(\tilde{x}) &=
\begin{cases}
    n_{\mathrm{c},i}, & \lvert \tilde{x} \rvert \leq D/2,\\
    n_{\mathrm{cl},i}, & \lvert \tilde{x} \rvert > D/2,
\end{cases}
\quad i=\{0,2,4\},
\label{eq:2}
\end{align}
where $D$ is the physical core width.

The proposed structure can be implemented with highly nonlinear core materials such as chalcogenide glasses ($\mathrm{As_2Se_3}$, $\mathrm{As_2S_3}$, or GLS), nanocomposites, or doped optical glasses \cite{Eggleton2011}. Combined with lower-index cladding layers of weaker nonlinear response, these materials can in principle provide the qualitative index and nonlinear contrasts assumed here using established integrated-photonics fabrication techniques \cite{Vien-waveguide,Hu:07,Xia-review}.

To normalize Eq.~(\ref{eq:1}), we introduce the dimensionless variables
$
\mathbf{r} =(x,y)
$
with
$
\mathbf{r}=\tilde{\mathbf{r}}/D, z = \tilde{z}/(k_0 n_{\mathrm{cl},0} D^2), 
$
and the complex field amplitude
$
\Psi = A\, k_0 D \sqrt{n_{\mathrm{cl},0} n_{\mathrm{c},2}}.
$
In dimensionless form, the spatial distributions of $n_0$, $n_2$, and $n_4$ become
\begin{align}
n_i(x) &=
\begin{cases}
    n_{\mathrm{c},i}, & \lvert x \rvert \leq 1/2,\\
    n_{\mathrm{cl},i}, & \lvert x \rvert > 1/2,
\end{cases}
\quad i=\{0,2,4\}.
\label{eq:3}
\end{align}

For convenience, define $\gamma(x)=n_2(x)/n_{\mathrm{c},2}$, $\sigma(x)=-n_4(x)/\big(n_{\mathrm{cl},0}n^2_{\mathrm{c},2}D^2k_0^2\big)$, and $U(x)=-n_{\mathrm{cl},0}D^2k_0^2\delta n_0(x)$. The corresponding linear and nonlinear profiles are
\begin{subequations}\label{eq:normalized_coefficients}
\begin{align}
U(x) &=
\begin{cases}
-\alpha=-n_{\mathrm{cl},0}(n_{\mathrm{c},0}-n_{\mathrm{cl},0})D^2k_0^2,
& \lvert x\rvert \leq 1/2,\\
0,
& \lvert x\rvert > 1/2,
\end{cases}
\\
\gamma(x) &=
\begin{cases}
1,
& \lvert x\rvert \leq 1/2,\\
\gamma_\mathrm{cl} = n_{\mathrm{cl},2}/n_{\mathrm{c},2},
& \lvert x\rvert > 1/2,
\end{cases}
\\
\sigma(x) &=
\begin{cases}
\sigma_\mathrm{c} = -n_{c,4}/\big(n_{\mathrm{cl},0} n_{\mathrm{c},2}^2 D^2 k_0^2 \big),
& \lvert x\rvert \leq 1/2,\\
\sigma_\mathrm{cl} = -n_{\mathrm{cl},4}/\big(n_{\mathrm{cl},0} n_{\mathrm{c},2}^2 D^2 k_0^2 \big),
& \lvert x\rvert > 1/2.
\end{cases}
\end{align}
\end{subequations}
The normalized nonlinear Schrodinger equation is therefore
\begin{equation}
    i\frac{\partial\Psi}{\partial z}=-\frac{1}{2}\Delta\Psi+U(x)\Psi-\gamma(x)|\Psi|^2\Psi+\sigma(x)|\Psi|^4\Psi,
    \label{eq:5}
\end{equation}
where $\Delta=\partial^2/\partial x^2+\partial^2/\partial y^2$. The model is specified by the independent parameters $\alpha$, $\gamma_\mathrm{cl}$, $\sigma_\mathrm{c}$, and $\sigma_\mathrm{cl}$. We focus on waveguides with a strong fifth-order response in the core and a weak one in the cladding, and therefore take $\sigma_\mathrm{cl}$ to be small in the calculations below. The conserved beam power is
\begin{equation}
    N=\int_{-\infty}^{\infty}d^2\mathbf{r}\ |\Psi(\mathbf{r},z)|^2.
    \label{eq:6}
\end{equation}
Stationary solutions of Eq.~(\ref{eq:5}) are sought as
$
\Psi(\mathbf{r},z) = \phi(\mathbf{r})\cdot\exp\!\left(i\mu z\right),
$ where $\mu$ is the propagation constant and $\phi(\mathbf{r})$ is real. The stationary equation is
\begin{equation}
    \mu \phi=\frac{1}{2}\nabla^2\phi 
    -U(x)\phi + \gamma(x)\phi^3-\sigma(x)\phi^5,
    \label{eq:7}
\end{equation}
with the propagation constant given by
\begin{equation}
    \mu=\frac{-1}{N}\int_{-\infty}^{\infty}d^2\mathbf{r}
    \bigg\{
        \frac{(\nabla\phi)^2}{2}+U(x)\phi^2-\gamma(x)\phi^4+\sigma(x)\phi^6
    \bigg\}.
    \label{eq:8}
\end{equation}

The form of the characteristic curve $N(\mu)$ depends strongly on the waveguide parameters. We analyze these curves below.

\section{Single-soliton solutions and their stability}
\subsection{Variational approximation}
For an approximate analytical description of the localized states of Eq.~(\ref{eq:5}), we use the VA. The corresponding Hamiltonian is
\begin{equation}
    \mathscr{H}
    =
    \int_{-\infty}^{\infty} d^2\mathbf{r} 
    \left\{
        \frac{(\nabla\phi)^2}{2}
        + U(x)\phi^2
        -\frac{\gamma(x)}{2}\phi^4
        +\frac{\sigma(x)}{3}\phi^6
    \right\}
    .
    \label{eq:9}
\end{equation}
We adopt the ansatz
\begin{equation}
    \phi(\mathbf{r})=A\cdot \exp{\left[-\left(\frac{x}{W}\right)^{2}\right]}  \exp{\left[-\left(\frac{y}{V}\right)^{2n}\right]}
    .
    \label{eq:10}
\end{equation}
Here $A$ is the amplitude, $W$ and $V$ are the widths along $x$ and $y$, and $n$ is the Gaussian order. Equation~(\ref{eq:6}) then gives
\begin{align}
N=\int_{-\infty}^\infty d^2\mathbf{r}\phi^2=A^2WV~2^{\frac{-1+n}{2n}}\sqrt{\pi}~\Gamma\left(1+\frac{1}{2n}\right)   
.
\label{eq:11}
\end{align}
Thus $A$ can be eliminated in favor of $N$, $W$, and $V$. Since $N$ is conserved, the variational problem reduces to the two widths $W$ and $V$. Substitution of Eq.~(\ref{eq:10}) into Eq.~(\ref{eq:9}) yields
\begin{align}
\mathscr{H}
&=
\frac{N}{2W^2}
+\frac{
2^{-2+\frac{1}{n}}\,n\,\Gamma\!\left(2-\frac{1}{2n}\right)N
}{
V^2\,\Gamma\!\left(1+\frac{1}{2n}\right)
}
-N\alpha
\operatorname{erf}\!\frac{1}{\sqrt{2}\,W}
\notag
\\
&-
\frac{N^2}{
4WV\sqrt{\pi}\,
\Gamma\!\left(1+\frac{1}{2n}\right)
}
\bigg[
(1-\gamma_\mathrm{cl})\operatorname{erf}\!\frac{1}{W}
+
\gamma_\mathrm{cl}
\bigg]
\notag
\\
&
+
\frac{
2^{-1+\frac{1}{n}}\,
3^{-\frac{3}{2}-\frac{1}{2n}}\,
N^3
}{
W^2 V^2 \pi\,
\Gamma^2\!\left(1+\frac{1}{2n}\right)
}
\bigg[
(\sigma_\mathrm{c}-\sigma_\mathrm{cl})
\operatorname{erf}\!\frac{\sqrt{3}}{\sqrt{2}W}
+\sigma_\mathrm{cl}
\bigg],
\label{eq:12}
\end{align}
where $\operatorname{erf}$ and $\Gamma$ denote the error and Gamma functions,
\begin{align}
    \operatorname{erf}x&=\frac{2}{\sqrt{\pi}}\int_0^xe^{-\zeta^2}d\zeta,
    \\
    \Gamma (x)&=\int_0^\infty \zeta^{x-1}e^{-\zeta}d\zeta.
\end{align}
Stationary variational states satisfy $\partial_W\mathscr{H}=\partial_V\mathscr{H}=0$, i.e.,
\begin{align}
0
=&
-\sqrt{\pi}\,W^2
+
\sqrt{2}\,e^{-\frac{1}{2W^2}}\alpha  W^3
+
\frac{N}
{4\,
\Gamma\!\left(1+\frac{1}{2n}\right)}
\notag\\
&\times\left[
\frac{4(1-\gamma_\mathrm{cl})}
{W\sqrt{\pi}\,e^{\frac{1}{W^2}}}
+
(1-\gamma_\mathrm{cl})\operatorname{erf}\!\frac{1}{W}
+
\gamma_\mathrm{cl}
\right]
-
\frac{
2^{\frac{1}{n}}
3^{-\frac{3}{2}-\frac{1}{2n}}
N^2
}{
V^2\Gamma^2\!\left(1+\frac{1}{2n}\right)
}
\notag
\\
&\times
\left[
\frac{
\sqrt{3}(\sigma_\mathrm{c}-\sigma_\mathrm{cl})
}{
W\sqrt{2\pi}e^{\frac{3}{2W^2}}
}
+
(\sigma_\mathrm{c}-\sigma_\mathrm{cl})
\operatorname{erf}\!\frac{\sqrt{3}}{\sqrt{2}W}
+\sigma_\mathrm{cl}
\right],
\label{eq:16}
\\
0
={}&
\frac{n\pi}{2}\,
\Gamma\!\left(2-\frac{1}{2n}\right)
-
\frac{\sqrt{\pi}\,VN}{2^{2+\frac{1}{n}}W}
\left[
(1-\gamma_\mathrm{cl})\operatorname{erf}\!\frac{1}{W}
+
\gamma_\mathrm{cl}
\right]
\notag
\\
&+
\frac{
3^{-\frac{3}{2}-\frac{1}{2n}}
N^2
}{
W^2
\Gamma\!\left(1+\frac{1}{2n}\right)
}
\left[
(\sigma_\mathrm{c}-\sigma_\mathrm{cl})
\operatorname{erf}\!\frac{\sqrt{3}}{\sqrt{2}W}
+\sigma_\mathrm{cl}
\right].
\label{eq:17}
\end{align}
The corresponding propagation constant is
\begin{align}
\mu
=&
-\frac{1}{2W^2}
-\frac{
2^{-2+\frac{1}{n}}\,n\,\Gamma\!\left(2-\frac{1}{2n}\right)
}{
V^2\,\Gamma\!\left(1+\frac{1}{2n}\right)
}
+\alpha
\operatorname{erf}\!\frac{1}{\sqrt{2}\,W}
\notag
\\
&+\frac{N}{
2W V\sqrt{\pi}\,
\Gamma\!\left(1+\frac{1}{2n}\right)
}
\left[
(1-\gamma_\mathrm{cl})\operatorname{erf}\!\frac{1}{W}
+
\gamma_\mathrm{cl}
\right]
\notag
\\
&-
\frac{
2^{-1+\frac{1}{n}}\,
3^{-\frac{1}{2}-\frac{1}{2n}}\,
N^2
}{
W^2 V^2\pi\,
\Gamma^2\left(1+\frac{1}{2n}\right)
}
\left[
(\sigma_\mathrm{c}-\sigma_\mathrm{cl})
\operatorname{erf}\!\frac{\sqrt{3}}{\sqrt{2}W}
+\sigma_\mathrm{cl}
\right].
\label{eq:15}
\end{align}
Equations~(\ref{eq:16})--(\ref{eq:17}) are solved numerically. Once $W$ and $V$ are obtained, stability within the VA is assessed from the Hessian matrix \cite{PhysRevLett.10.1103}, $K_{ij}=\partial^2\mathscr{H}/\partial q_i\partial q_j$, with $q=\{W,V\}$. A positive-definite Hessian, i.e., positive eigenvalues, identifies a local minimum of the Hamiltonian and hence a stable variational state; a negative eigenvalue signals instability.

The VA predictions are compared below with numerical solutions of Eq.~(\ref{eq:5}).

\begin{figure*}[!tb]
    \centering
    \includegraphics[width=\linewidth]{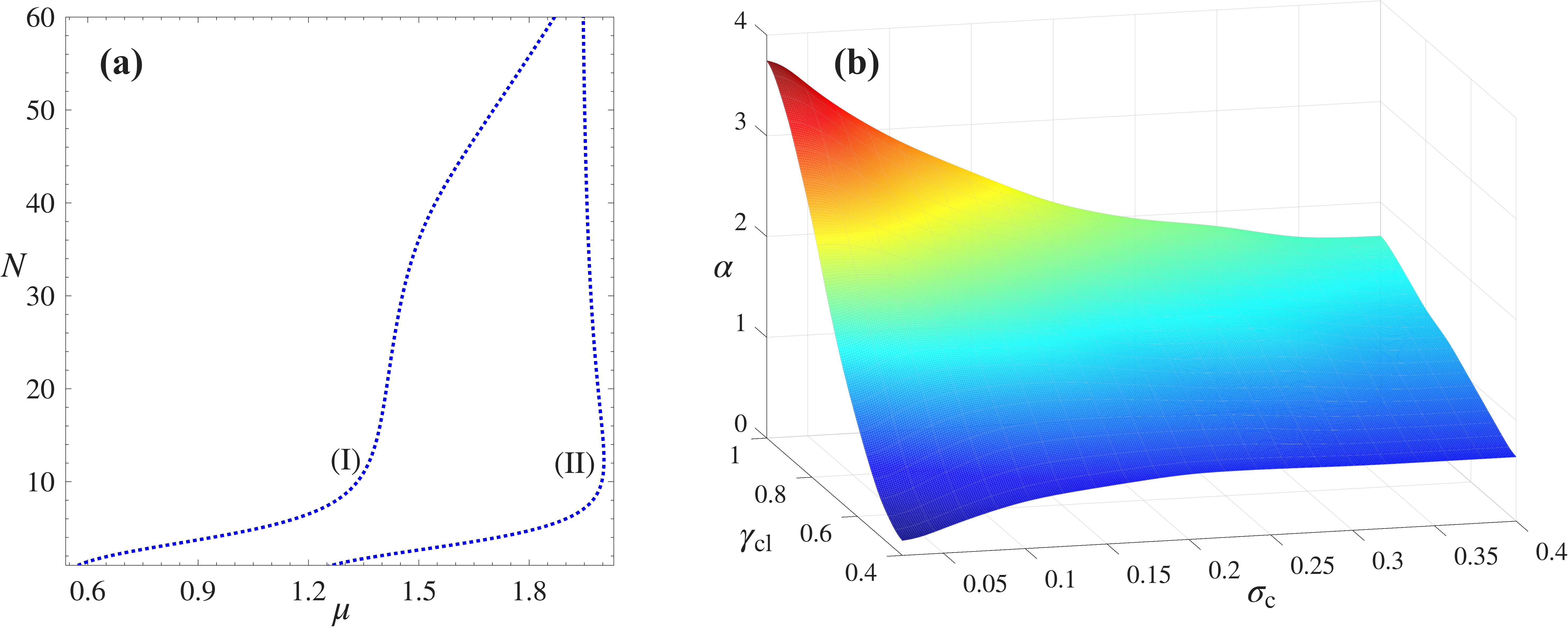}
    \caption{(a) VA characteristic curves $N(\mu)$ for $n=1$, $\gamma_\mathrm{cl}=0.98$, $\sigma_\mathrm{c}=0.2$, and $\sigma_\mathrm{cl}=0.01$: $\alpha=1.5$ gives type~(I), whereas $\alpha=2.0$ gives type~(II). (b) Boundary surface predicted by the VA in $(\sigma_\mathrm{c},\gamma_\mathrm{cl},\alpha)$ space for $\sigma_\mathrm{cl}=0.01$, separating the parameter regions of type-(I) and type-(II) curves.}
    \label{fig:2}
\end{figure*}

\subsection{Numerical methods and linear stability analysis}

In addition to the VA, stationary solitons are computed numerically. We combine AITM, SOM, and NCGM \cite{yang2010nonlinear} to trace the solutions in different parameter regions, using nearby VA predictions as initial estimates.

Their stability is first checked by direct split-step Fourier propagation after adding a small perturbation to the initial stationary state \cite{yang2010nonlinear}.

We also apply linear stability analysis (LSA) \cite{yang2010nonlinear} to the stationary solutions of Eq.~(\ref{eq:7}). A small, generally complex perturbation is introduced as
\begin{align}
    \psi(\mathbf{r},z)=
    \left[
        \phi(\mathbf{r})+\delta\psi(\mathbf{r},z)
    \right]
    e^{i\mu z},
    \label{eq:18}
\end{align}
where $|\delta\psi(\mathbf{r},z)|\ll|\psi(\mathbf{r},z)|$ and $\phi(\mathbf{r})$ is real. Substitution into the governing Eq.~(\ref{eq:5}) and linearization give
\begin{align}
    i\partial_z\delta\psi
    =&
    \left[-\nabla^2/2+\mu+U(x)-2\gamma(x)\phi^2+3\sigma(x)\phi^4 \right]\delta\psi
    \notag
    \\
    &+
    \left[
        -\gamma(x)+2\sigma(x)\phi^2 
    \right]
    \phi^2\delta\psi^\ast
    .
    \label{eq:19}
\end{align}
This equation can be written in matrix form as
\begin{gather}
i\partial_z
\begin{pmatrix}
\delta\psi \\
\delta\psi^\ast
\end{pmatrix}
=
\begin{pmatrix}
\widehat{A} & \widehat{B} \\
-\widehat{B} & -\widehat{A}
\end{pmatrix}
\begin{pmatrix}
\delta\psi \\
\delta\psi^\ast
\end{pmatrix},
\label{eq:20}
\\
\widehat{A}
\equiv
-(1/2)\nabla^2+\mu + U(x)
- 2\gamma(x) \phi^2
+ 3\sigma(x) \phi^4,
\label{eq:21}
\\
\widehat{B}
\equiv
-\gamma(x)\phi^2
+ 2\sigma(x)\phi^4.
\label{eq:22}
\end{gather}
We further set
\begin{align}
    \delta\psi(\mathbf{r},z)=f(\mathbf{r})e^{\lambda z}+g^\ast(\mathbf{r})e^{\lambda^\ast z},
    \label{eq:23}
\end{align}
where $\lambda$ is the complex eigenvalue and $f$ and $g$ are complex eigenfunctions. Its real part gives the exponential growth rate. Substitution into Eq.~(\ref{eq:20}) yields
\begin{equation}
-i
\begin{pmatrix}
0 & \widehat{C} \\
\widehat{D} & 0
\end{pmatrix}
\begin{pmatrix}
\eta \\
\xi
\end{pmatrix}
=
\lambda
\begin{pmatrix}
\eta \\
\xi
\end{pmatrix}.
\label{eq:24}
\end{equation}
where $\xi=f(\mathbf{r})+g(\mathbf{r})$, $\eta=f(\mathbf{r})-g(\mathbf{r})$, $\widehat{C}=\widehat{A}+\widehat{B}$, and $\widehat{D}=\widehat{A}-\widehat{B}$.

Equation~(\ref{eq:24}) is solved by Fourier collocation \cite{yang2010nonlinear}. A stationary state is unstable if any eigenvalue has $\operatorname{Re}(\lambda)>0$; if all eigenvalues satisfy $\operatorname{Re}(\lambda)=0$, no perturbation mode grows exponentially and the state is linearly stable.

\subsection{Numerical results and comparison with analytical approximation}

We now compare the VA with numerical solutions of Eq.~\eqref{eq:5}, focusing on the $N(\mu)$ curves, field profiles, and stability.

We first survey the Gaussian VA ($n=1$) over a broad parameter range. Its characteristic curves $N(\mu)$ fall into two qualitatively different types, labeled (I) and (II) in Fig.~\ref{fig:2}(a). Type~(I) is monotonic, with $N$ increasing with $\mu$. Type~(II) becomes nonmonotonic at high power: $\mu$ reaches a maximum and then decreases as $N$ grows further.

\begin{figure*}[!tb]
    \centering
    \includegraphics[width=\linewidth]{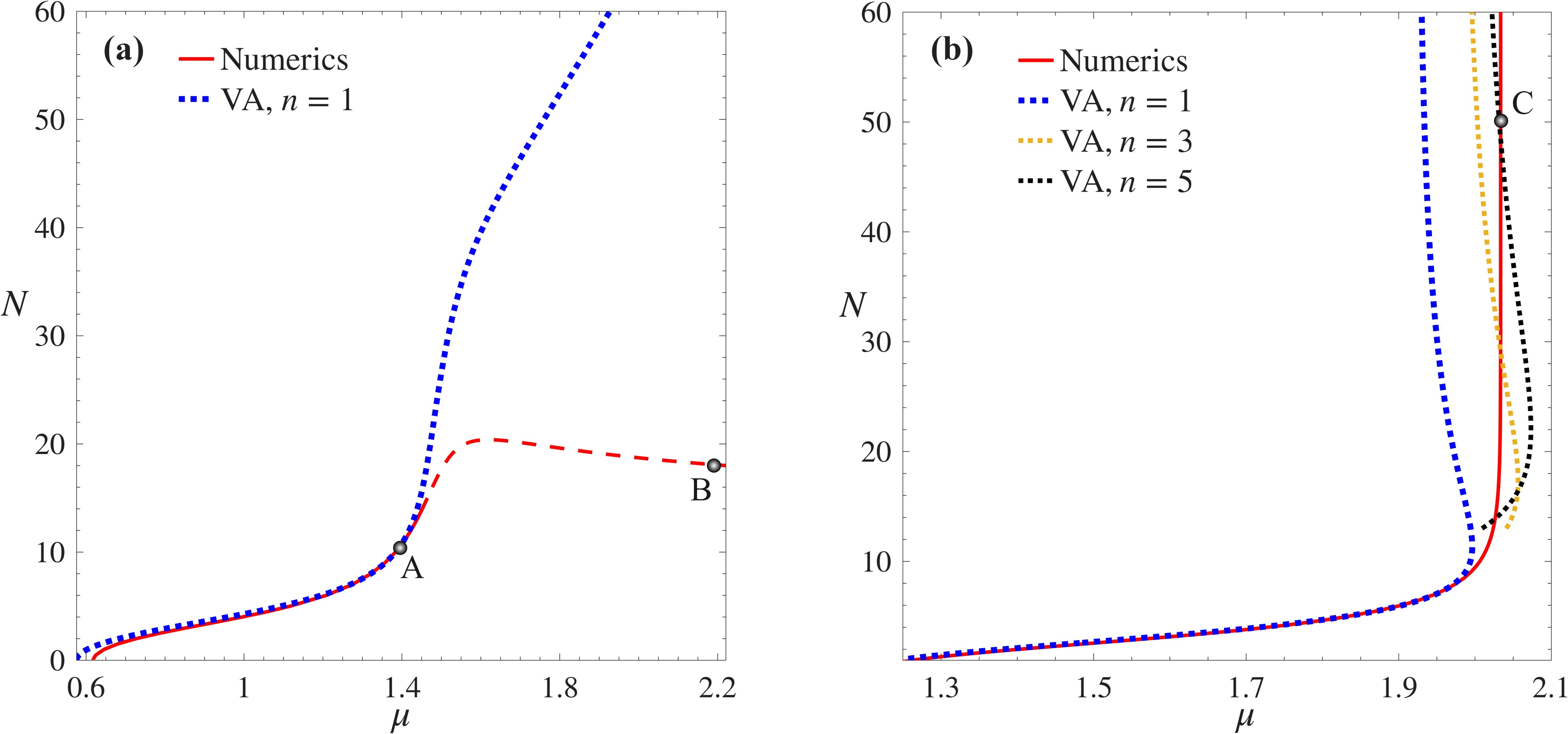} 
    \caption{(a) $N(\mu)$ from the VA ($n=1$) and numerical calculations for the type-(I) parameter set of Fig.~\ref{fig:2}(a), with $\alpha=1.5$. Blue dotted: VA; red: numerics (solid, stable; dashed, unstable). (b) $N(\mu)$ from the VA ($n=1,3,5$) and numerics for the type-(II) parameter set, with $\alpha=2$. All states shown in panel (b) are stable.}
    \label{fig:3}
\end{figure*}
\begin{figure*}[!tb]
    \centering
    \includegraphics[width=\linewidth]{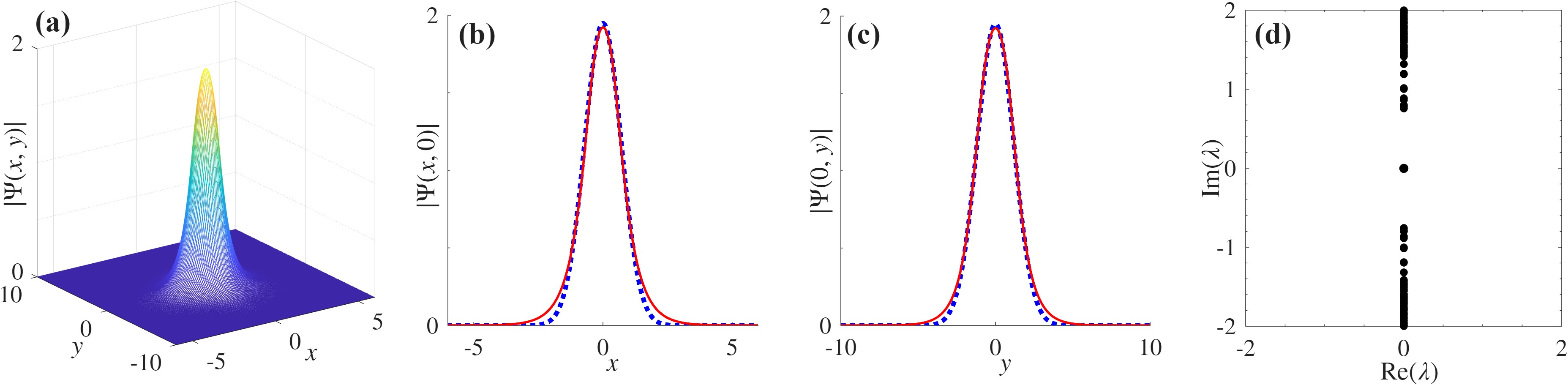}
    \caption{Stable soliton at point $A$ ($N=10$) in Fig.~\ref{fig:3}(a). (a) Numerical three-dimensional profile. (b,c) Cross sections at $y=0$ and $x=0$, respectively: VA ($n=1$, blue dotted) and numerical solution (red solid). (d) Linear-stability spectrum.}
    \label{fig:4}
\end{figure*}

Varying $(\sigma_{\mathrm{c}},\gamma_{\mathrm{cl}},\alpha)$ at fixed small $\sigma_{\mathrm{cl}}$, the Gaussian VA yields the boundary surface shown in Fig.~\ref{fig:2}(b). Parameter sets below this surface generate type-(I) curves, whereas those above it generate type-(II) curves. The Hessian test classifies all VA states obtained in this survey as stable.

We then solve Eq.~\eqref{eq:5} numerically for representative parameter sets on both sides of the surface. The numerical $N(\mu)$ curves display the same two qualitative types, but the accuracy of the VA differs between the two regions. Below the surface, substantial qualitative deviations develop at large $N$ and $\mu$; above it, the variational and numerical curves retain the same overall form. Thus, although obtained solely from the Gaussian VA, the surface also provides a useful classification of the characteristic curves of the full numerical problem.

\begin{figure*}[!tb]
    \centering
    \includegraphics[width=\linewidth]{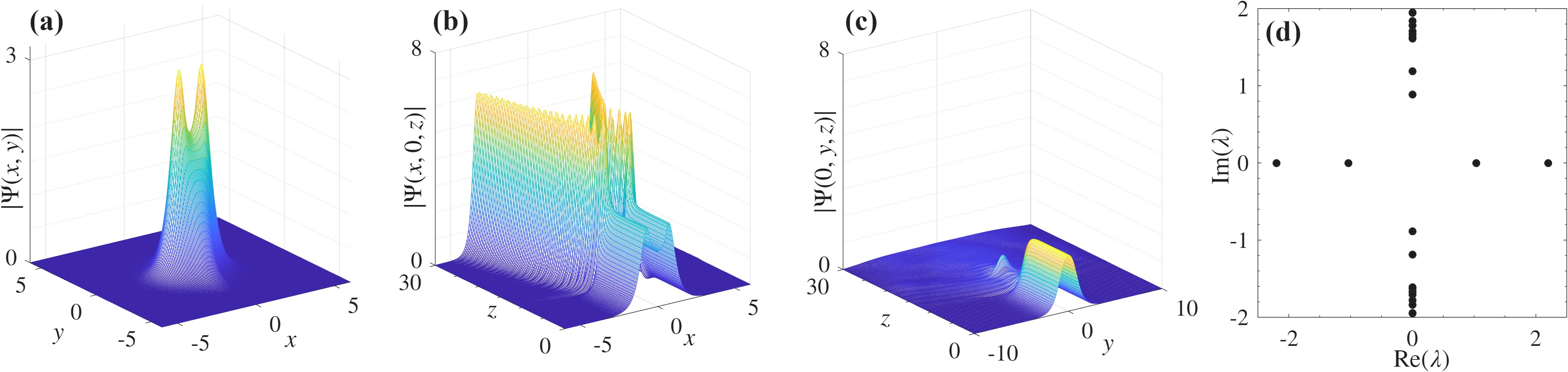}
    \caption{Unstable soliton at point $B$ ($N=18.2$) in Fig.~\ref{fig:3}(a). (a) Numerical three-dimensional profile. (b,c) Evolution of cross sections at $y=0$ and $x=0$, respectively. (d) Linear-stability spectrum.}
    \label{fig:5}
\end{figure*}

As an example below the boundary surface, we take the type-(I) set with $\alpha=1.5$, shown in Fig.~\ref{fig:3}(a). At low power, the Gaussian VA closely follows the numerical branch. The numerical soliton is stable, single-peaked, and nearly Gaussian; point $A$ in Fig.~\ref{fig:4} illustrates the agreement in both cross sections, while the linear-stability spectrum and direct propagation confirm its stability. With increasing power, however, once $N>16$ and $\mu>1.48$, the numerical profile deforms along the confined $Ox$ direction and develops two symmetric peaks. The numerical $N(\mu)$ curve reaches $N_{\max}\approx20.4$ at $\mu\approx1.6$ and then bends backward. These two-peak states are linearly unstable. Point $B$ in Fig.~\ref{fig:5} shows a representative profile and its instability; direct propagation further shows breakup into secondary beams that leave the guiding channel. Hence, below the surface the Gaussian VA is accurate at low power but fails to reproduce the high-power profile, stability, and characteristic curve.

Above the boundary surface we choose the type-(II) set with $\alpha=2$, Fig.~\ref{fig:3}(b). Here the numerical and VA characteristic curves have the same qualitative form. At low power the numerical state is nearly Gaussian and is reproduced closely by the Gaussian VA, while LSA finds the numerical branch stable over the range examined. For $N>17$, the state broadens mainly along the free $Oy$ direction and gradually develops a flat top. The Gaussian VA still captures the qualitative $N(\mu)$ behavior, although its profile becomes less accurate as the flat-top character strengthens.

\begin{figure*}[!tb]
    \centering
    \includegraphics[width=\linewidth]{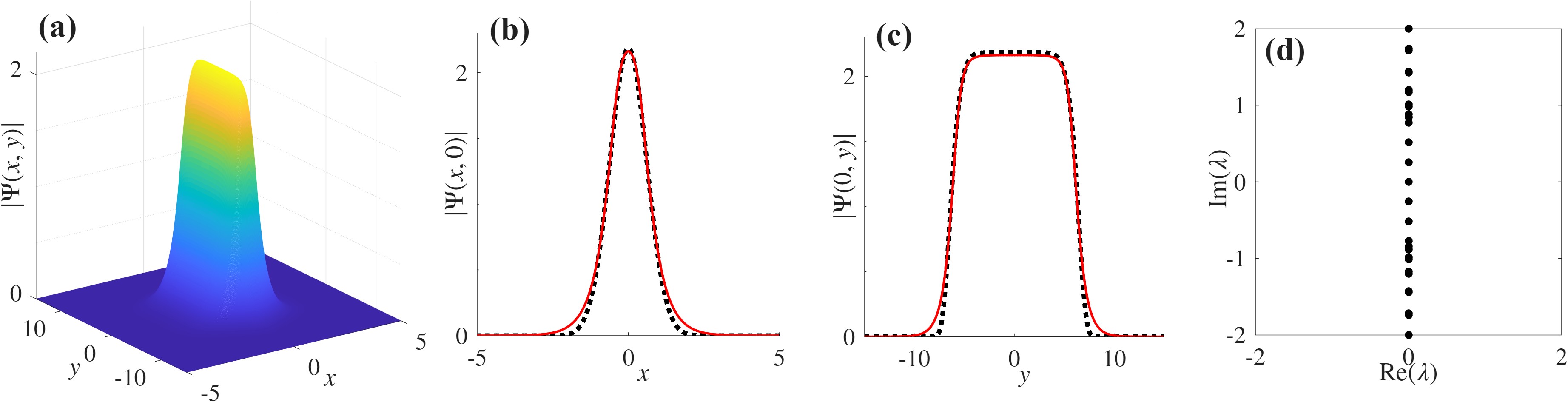} 
    \caption{Stable soliton at point $C$ ($N=50$) in Fig.~\ref{fig:3}(b). (a) Numerical three-dimensional profile. (b,c) Cross sections at $y=0$ and $x=0$, respectively: VA ($n=5$, blue dotted) and numerical solution (red solid). (d) Linear-stability spectrum.}
    \label{fig:6}
\end{figure*}

The high-power type-(II) states are described more accurately by a super-Gaussian ansatz ($n>1$). In Fig.~\ref{fig:3}(b), the $n=3$ and $n=5$ curves approach the numerical result much more closely than the Gaussian ($n=1$) prediction. Point $C$ ($N=50$), shown in Fig.~\ref{fig:6}, confirms this improvement at the profile level. The super-Gaussian ansatz reproduces the broad flat-top region along $Oy$ while retaining confinement along $Ox$.


Other parameter sets on both sides of the surface show the same qualitative behavior. The potential depth $\alpha$, equivalently the linear-index contrast between core and cladding, is particularly important for the classification into the two $N(\mu)$ types, whereas $\sigma_c$ strongly influences the formation of high-power flat-top states. This behavior differs from the pure Kerr case, where self-focusing is bounded by the Townes-soliton threshold.

\section{Nonlinear dynamics of soliton interaction}
We next study nonlinear interactions of the solitons obtained above, comparing the HVA with direct numerical simulations.

\subsection{Hybrid variational approximation}
Because $U(x)$, $\gamma(x)$, and $\sigma(x)$ depend only on the confined coordinate $x$, the system remains translationally invariant along the free direction $y$. This geometry is well suited to the HVA: the $x$ dependence is represented variationally, whereas the dynamics along $y$ remain functional. The original two-dimensional problem is thereby reduced to an effective evolution system that retains the essential interaction dynamics.

We use the HVA ansatz
\begin{equation}
    \Psi(\mathbf{r},z)=\frac{\psi(y,z)}{(\pi/2)^{\frac{1}{4}}W(z)^{\frac{1}{2}}}\cdot\exp{\left[-\frac{x^2}{W^2(z)}+i b(z)~x^2\right]},
    \label{eq:25}
\end{equation}
where $W(z)$ and $b(z)$ are the width and spatial chirp along the confined $x$ direction. Substitution into Eq.~\eqref{eq:6} gives
\begin{equation}
    N=\int_{-\infty}^{\infty}d^2\mathbf{r}|\Psi(\mathbf{r},z)|^2=\int_{-\infty}^{\infty}dy|\psi(z,y)|^2.
    \label{eq:26}
\end{equation}
The variational quantities $\xi=\{W,b,\psi\}$, with $\xi_y=\partial\xi/\partial y$ and $\xi_z=\partial\xi/\partial z$, obey
\begin{equation}
    \frac{\partial \mathscr{L}_H}{\partial\xi^*}
    -
    \frac{\partial}{\partial y}\frac{\partial \mathscr{L}_H}{\partial\xi^*_y}
    -
    \frac{\partial}{\partial z}\frac{\partial \mathscr{L}_H}{\partial\xi^*_z}=0,
    \label{eq:28}
\end{equation}
where $\mathscr{L}_H$ is the hybrid Lagrangian,
\begin{align}
    \mathscr{L}_H=\int_{-\infty}^{\infty}dx\bigg\{&\frac{i}{2}\bigg(\Psi\frac{\partial\Psi^*}{\partial z}-\Psi^*\frac{\partial\Psi}{\partial z}\bigg)+\frac{|\nabla\Psi|^2}{2}+U(x)|\Psi|^2
    \notag\\&-\frac{\gamma(x)}{2}|\Psi|^4+\frac{\sigma(x)}{3}|\Psi|^6\bigg\}.
    \label{eq:41}
\end{align}

\begin{figure*}[!tb]
    \centering
    \includegraphics[width=\linewidth]{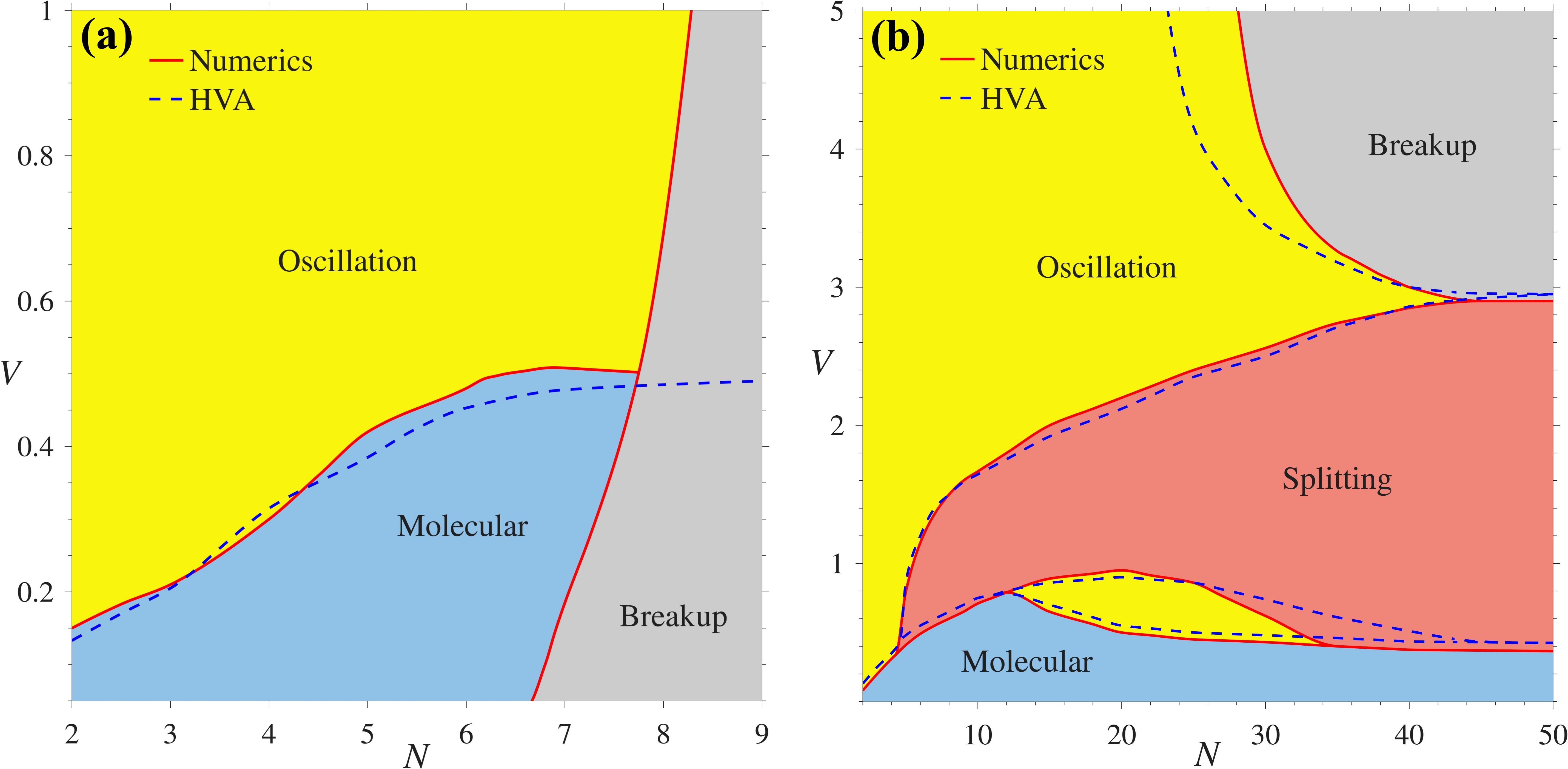}
    \caption{Post-collision dynamical states in the $(N,v)$ plane. (a) Parameter set of Fig.~\ref{fig:3}(a): numerics show Oscillation, Molecular, and Breakup states, while the HVA reproduces Oscillation and Molecular. (b) Parameter set of Fig.~\ref{fig:3}(b): both approaches identify Oscillation, Molecular, Splitting, and Breakup. The Oscillation region at $N<18$ corresponds to nearly elastic collisions, whereas the high-power Oscillation regime is inelastic.}
    \label{fig:7}
\end{figure*}

Substitution of Eq.~\eqref{eq:25} into Eq.~\eqref{eq:41} gives
\begin{align}
        \mathscr{L}_H
        =&
        \frac{|\psi_y|^2}{2}
        +
        \frac{i}{2}(\psi\psi^\ast_z
        -
        \psi^\ast\psi_z)
        -
        \frac{|\psi|^4}{2\sqrt{\pi}W}
        \bigg[
        (1-\gamma_\mathrm{cl})\operatorname{erf}\!\frac{1}{W}
        +
        \gamma_\mathrm{cl}
        \bigg]
        \notag\\
        &+
        \frac{|\psi|^2}{2}
        \left[
            \frac{1}{W^2}-2\alpha\operatorname{erf}\frac{1}{\sqrt{2}W}+\left(b^2+\frac{b'}{2}\right)W^2
        \right]
        \notag\\
        &+
        \frac{2|\psi|^6}{3\sqrt{3}\pi W^2}
        \bigg[
            (\sigma_\mathrm{c}-\sigma_\mathrm{cl})
            \operatorname{erf}\!\frac{\sqrt{3}}{\sqrt{2}W}
            +\sigma_\mathrm{cl}
        \bigg].
    \label{eq:27}
\end{align}
Using Eq.~(\ref{eq:27}) in the Euler--Lagrange equation~(\ref{eq:28}) yields the effective equation for the $y$-dependent field,
\begin{align}
    i\frac{\partial\psi}{\partial z}
    =&
    -\frac{1}{2}\frac{\partial^2\psi}{\partial y^2}
    +
    \left[
        -\alpha\operatorname{erf}\frac{1}{\sqrt{2}W}+\frac{1}{2W^2}+\frac{b^2W^2}{2}+\frac{W^2}{4}\frac{db}{dz}
    \right]
    \psi
    \notag\\
    &-
    \left[
        (1-\gamma_\mathrm{cl})\operatorname{erf}\frac{1}{W}+\gamma_\mathrm{cl}
    \right]
    \frac{|\psi|^2\psi}{\sqrt{\pi}W}
    \notag\\
    &+
    \left[
        (\sigma_\mathrm{c}-\sigma_\mathrm{cl})\operatorname{erf}\frac{\sqrt{3}}{\sqrt{2}W}+\sigma_\mathrm{cl}
    \right]
    \frac{2|\psi|^4\psi}{\sqrt{3}\pi W^2}.
    \label{eq:29}
\end{align}
The chirp satisfies
\begin{align}
        b=\frac{1}{2W}\frac{dW}{dz},
        \label{eq:30}
\end{align}
and the width obeys
\begin{align}
    \frac{d^2W}{dz^2}
    =&
    \frac{4}{W^3}-\frac{4\sqrt{2}~e^{-\frac{1}{2W^2}}\alpha}{\sqrt{\pi}~W^2}
    -
    \frac{2}{\sqrt{\pi}~W^2}
    \bigg[
        \frac{2(1-\gamma_\mathrm{cl})}{\sqrt{\pi}We^{\frac{1}{W^2}}}
        +
        \gamma_\mathrm{cl}
        \notag\\
    &+
        (1-\gamma_\mathrm{cl})\operatorname{erf}\frac{1}{W}
    \bigg]
        |\psi|^2
        +
        \frac{16}{3\sqrt{3}\pi W^3}
    \bigg[
        \frac{\sqrt{3}(\sigma_\mathrm{c}-\sigma_\mathrm{cl})}{\sqrt{2\pi}We^{\frac{3}{2W^2}}}
        +
        \sigma_\mathrm{cl}
        \notag\\
        &+
        (\sigma_\mathrm{c}-\sigma_\mathrm{cl})\operatorname{erf}\frac{\sqrt{3}}{\sqrt{2}W}
    \bigg]
    |\psi|^4.
    \label{eq:31}
\end{align}
Multiplying Eq.~(\ref{eq:31}) by $|\psi|^2dy$ and integrating over $y$ gives
\begin{align}
\frac{d^2W}{dz^2}
    ={}&
    \frac{4}{W^3}-\frac{4\sqrt{2}~e^{-\frac{1}{2W^2}}\alpha}{\sqrt{\pi}~W^2}
    \notag\\
    &+
    \bigg[
        \frac{\sqrt{3}(\sigma_\mathrm{c}-\sigma_\mathrm{cl})}{\sqrt{2\pi}We^{\frac{3}{2W^2}}}
        +
        (\sigma_\mathrm{c}-\sigma_\mathrm{cl})\operatorname{erf}\frac{\sqrt{3}}{\sqrt{2}W}
        +
        \sigma_\mathrm{cl}
    \bigg]\mathcal{I}_2
    \notag\\
    &-
    \bigg[
        \frac{2(1-\gamma_\mathrm{cl})}{\sqrt{\pi}We^{\frac{1}{W^2}}}
        +
        (1-\gamma_\mathrm{cl})\operatorname{erf}\frac{1}{W}
        +
        \gamma_\mathrm{cl}
    \bigg]\mathcal{I}_1,
    \label{eq:32}
\end{align}
where
\begin{align}
    \mathcal{I}_1&=\frac{2}{\sqrt{\pi}~W^2N}\int_{-\infty}^{\infty}dy|\psi|^4,
    \\
    \mathcal{I}_2&=\frac{16\sqrt{3}}{9\pi W^3N}\int_{-\infty}^{\infty}dy|\psi|^6.
\end{align}
Following \cite{Edwards2005Hybrid}, we introduce
\begin{align}
    \Pi(z)=&-\alpha\operatorname{erf}\frac{1}{\sqrt{2}W}+\frac{1}{2W^2}+\frac{b^2W^2}{2}+\frac{W^2}{4}\frac{db}{dz}
    \notag\\
    =&
    -\alpha\operatorname{erf}\frac{1}{\sqrt{2}W}+\frac{1}{W^2}-\frac{\alpha e^{-\frac{1}{2W^2}}}{\sqrt{2\pi}W}
    \notag\\
    &+ \frac{W}{8}\bigg[
        \frac{\sqrt{3}(\sigma_\mathrm{c}-\sigma_\mathrm{cl})}{\sqrt{2\pi}We^{\frac{3}{2W^2}}}
        +
        (\sigma_\mathrm{c}-\sigma_\mathrm{cl})\operatorname{erf}\frac{\sqrt{3}}{\sqrt{2}W}
        +
        \sigma_\mathrm{cl}
    \bigg]\mathcal{I}_2
    \notag\\
    &-
    \frac{W}{8}\bigg[
        \frac{2(1-\gamma_\mathrm{cl})}{\sqrt{\pi}We^{\frac{1}{W^2}}}
        +
        (1-\gamma_\mathrm{cl})\operatorname{erf}\frac{1}{W}
        +
       \gamma_\mathrm{cl}
    \bigg]\mathcal{I}_1,    
\end{align}
together with
\begin{align}
    \Theta(z)=\int_0^zd\zeta~\Pi(\zeta),
    \\
    \tilde{\psi}(y,z)=\psi(y,z)e^{i\Theta(z)}.
\end{align}
Equation~(\ref{eq:29}) then reduces to
\begin{align}
    i\frac{\partial\tilde{\psi}}{\partial z}
    =&
    -\frac{1}{2}\frac{\partial^2\tilde{\psi}}{\partial y^2}
    -
    \left[
        (1-\gamma_\mathrm{cl})\operatorname{erf}\frac{1}{W}+\gamma_\mathrm{cl}
    \right]
    \frac{|\tilde{\psi}|^2\tilde{\psi}}{\sqrt{\pi}W}
    \notag\\
    &+
    \left[
        (\sigma_\mathrm{c}-\sigma_\mathrm{cl})\operatorname{erf}\frac{\sqrt{3}}{\sqrt{2}W}+\sigma_\mathrm{cl}
    \right]
    \frac{2|\tilde{\psi}|^4\tilde{\psi}}{\sqrt{3}\pi W^2}.
    \label{eq:38}
\end{align}

Thus, within the HVA, the interaction dynamics are governed by Eqs.~\eqref{eq:30}, \eqref{eq:32}, and \eqref{eq:38} for $b$, $W$, and $\tilde{\psi}$.

\subsection{Comparison between HVA and numerical simulations}

\begin{figure*}[!tb]
    \centering
    \includegraphics[width=\linewidth]{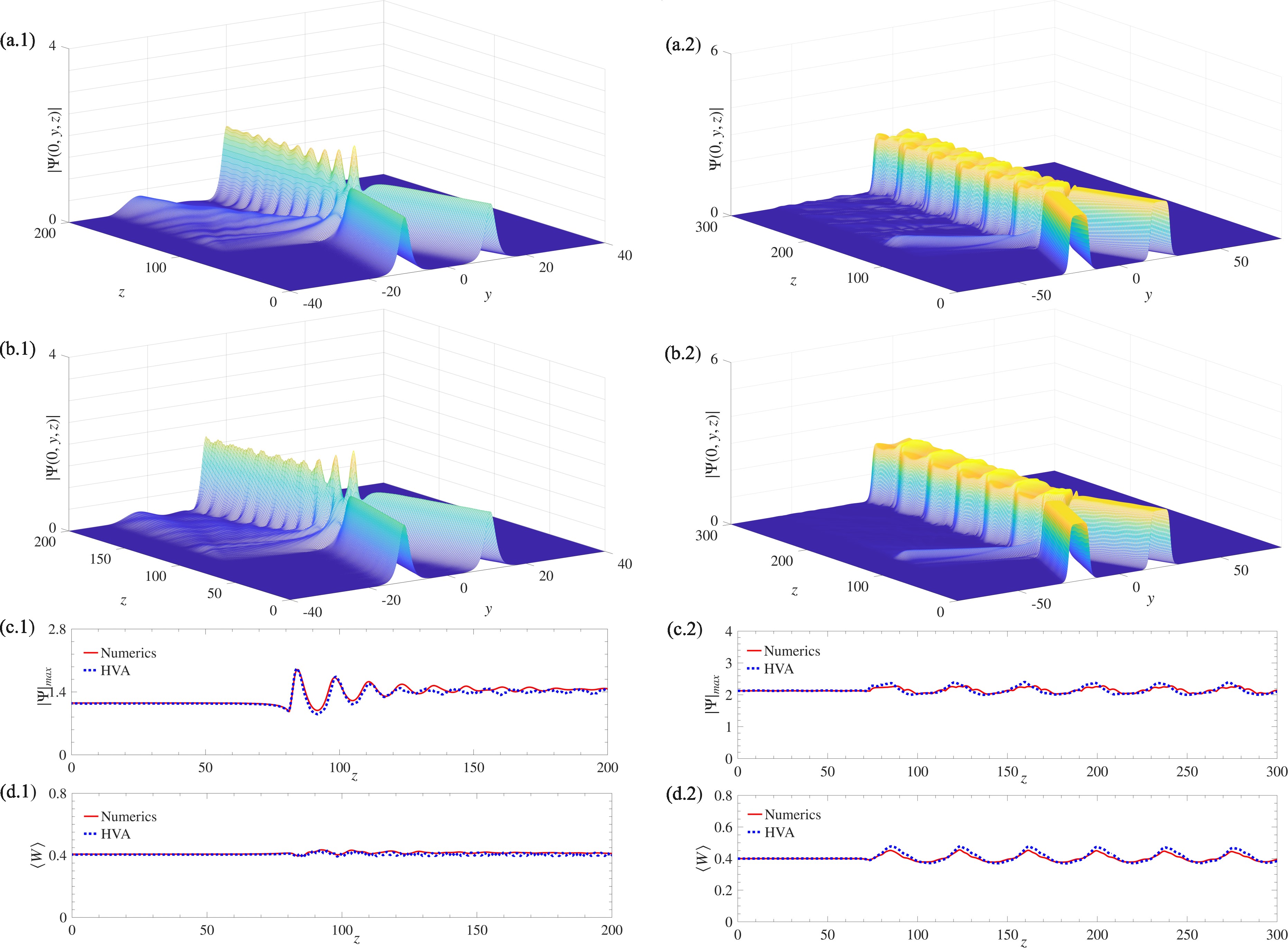}
    \caption{Molecular states. Left column: Fig.~\ref{fig:7}(a), $N=4$, $v=0.1$; right column: Fig.~\ref{fig:7}(b), $N=50$, $v=0.02$. Panels (a.1,b.1) and (a.2,b.2) show the $yz$ evolution of $|\Psi(\mathbf{r},z)|$ from direct simulations and the HVA, respectively ($n=1$ on the left and $n=5$ on the right). Panels (c.1,d.1) and (c.2,d.2) compare $|\Psi(\mathbf{r},z)|_{\max}$ and $\langle W(z)\rangle$.}
    \label{fig:8}
\end{figure*}

We collide two identical solitons with the same $N$ and $\mu$, assigning equal and opposite velocities $\pm v$. In the full simulations, stationary numerical solutions from Sec.~3.2 are multiplied by $\exp(\pm ivy)$ and propagated with the split-step Fourier method \cite{yang2010nonlinear,agrawal2019nonlinear}. The HVA starts from the corresponding VA state with the same phase factors. Before use as an HVA initial condition, each VA profile is checked against its numerical counterpart to ensure an adequate representation of the stationary soliton. We use the Gaussian ansatz ($n=1$) at low power and super-Gaussian forms ($n=2,3,\ldots$) at high power. Absorbing boundaries suppress reflections of radiation from the edges of the numerical domain.

During the interaction, the width along the confined $Ox$ direction may oscillate under the combined action of $U(x)$, $\gamma(x)$, and $\sigma(x)$. We characterize it by
\begin{align}
    \langle W(z)\rangle=\frac{\sqrt{\pi}}{\sqrt{2}N}\int_{-\infty}^{\infty}d^2\mathbf{r}|\Psi(\mathbf{r},z)|^2|x|.
\end{align}
We also monitor the peak amplitude $|\Psi(z)|_{\max}$. Together, these two quantities provide a convenient measure of the collision-induced deformation.

We first use a parameter set below the boundary surface in Fig.~\ref{fig:2}(b), corresponding to the type-(I) curve in Fig.~\ref{fig:2}(a). Scanning $N$ and $v$ gives the post-collision map in Fig.~\ref{fig:7}(a). At low power, the HVA reproduces both the observed states and their boundaries with good accuracy.

At small $N$ and $v$, the solitons do not separate after impact but merge into a single bound structure, which we denote the Molecular state; see the left column of Fig.~\ref{fig:8}. Before the collision, $\langle W(z)\rangle$ and $|\Psi(z)|_{\max}$ are nearly constant. The newly formed bound state is initially excited, producing a large but rapidly damped oscillation of $|\Psi(z)|_{\max}$, while $\langle W(z)\rangle$ changes only weakly and approaches a steady value. A fraction of the excess energy is carried away by radiation, allowing the localized field to relax toward the bound state. Thus the Molecular regime represents fusion into a ``soliton molecule'' that subsequently propagates as one object. The HVA reproduces both the formation of this state and the evolution of $|\Psi(z)|_{\max}$ and $\langle W(z)\rangle$ with very good accuracy.

\begin{figure*}[!tb]
    \centering
    \includegraphics[width=\linewidth]{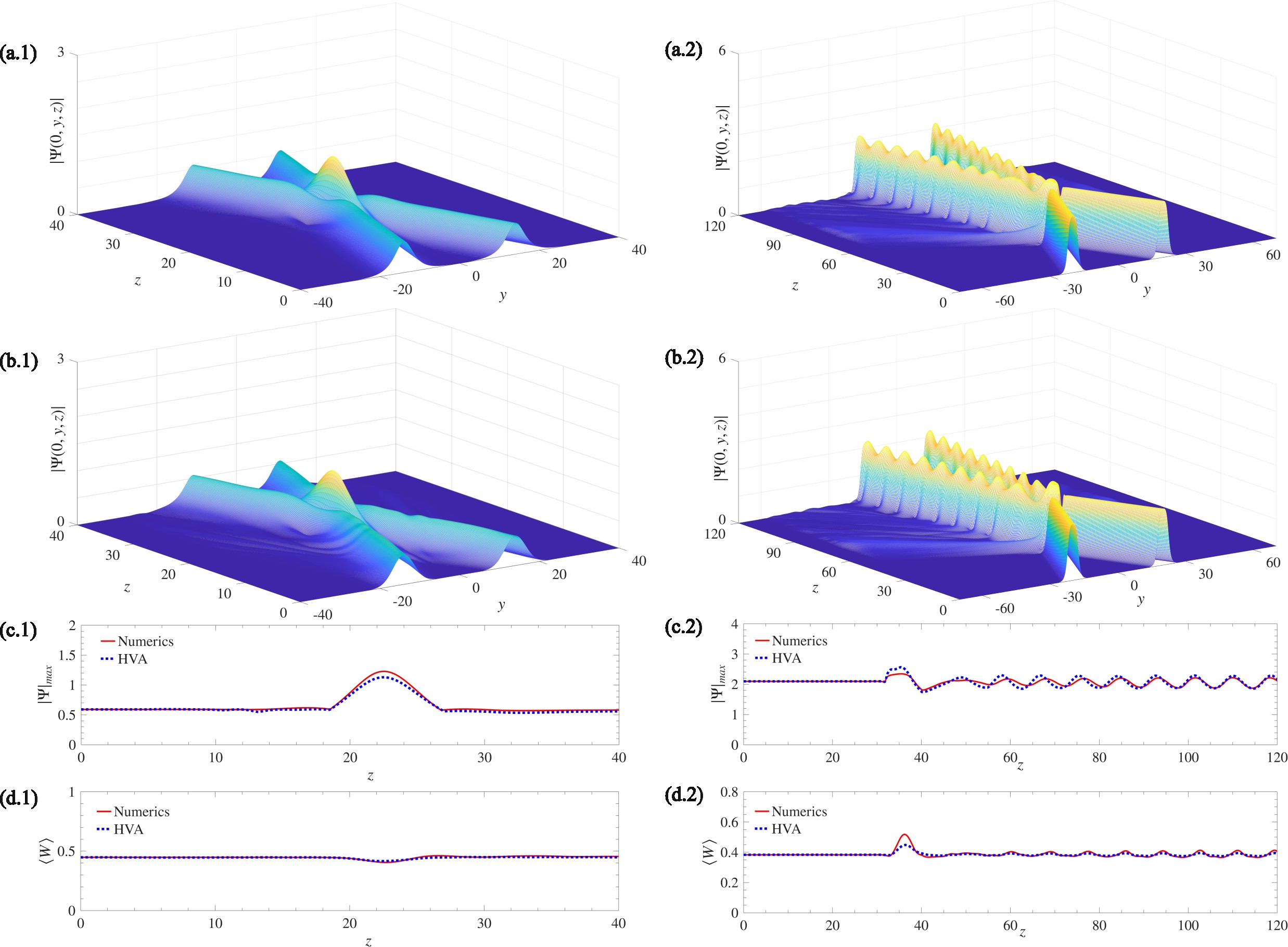}
    \caption{Oscillation states. Left column: Fig.~\ref{fig:7}(a), $N=2$, $v=0.6$; right column: Fig.~\ref{fig:7}(b), $N=20$, $v=0.5$. Panels (a.1,b.1) and (a.2,b.2) show the $yz$ evolution of $|\Psi(\mathbf{r},z)|$ from direct simulations and the HVA, respectively ($n=1$ on the left and $n=2$ on the right). Panels (c.1,d.1) and (c.2,d.2) compare $|\Psi(\mathbf{r},z)|_{\max}$ and $\langle W(z)\rangle$.}
    \label{fig:9}
\end{figure*}

At larger $v$, the outcome changes to the Oscillation state: the solitons separate after the collision and continue to propagate, but with oscillating amplitudes and widths. The left column of Fig.~\ref{fig:9} shows close agreement between the HVA and direct simulations. As the wavefunctions overlap, the peak amplitude rises sharply and is maximal near complete overlap, while the transverse width is also perturbed. After passage, both $|\Psi(z)|_{\max}$ and $\langle W(z)\rangle$ oscillate. Because only a small fraction of the soliton power is radiated, this regime is close to an elastic collision, apart from the post-collision excitation.

\begin{figure*}[!tb]
    \centering
    \includegraphics[width=\linewidth]{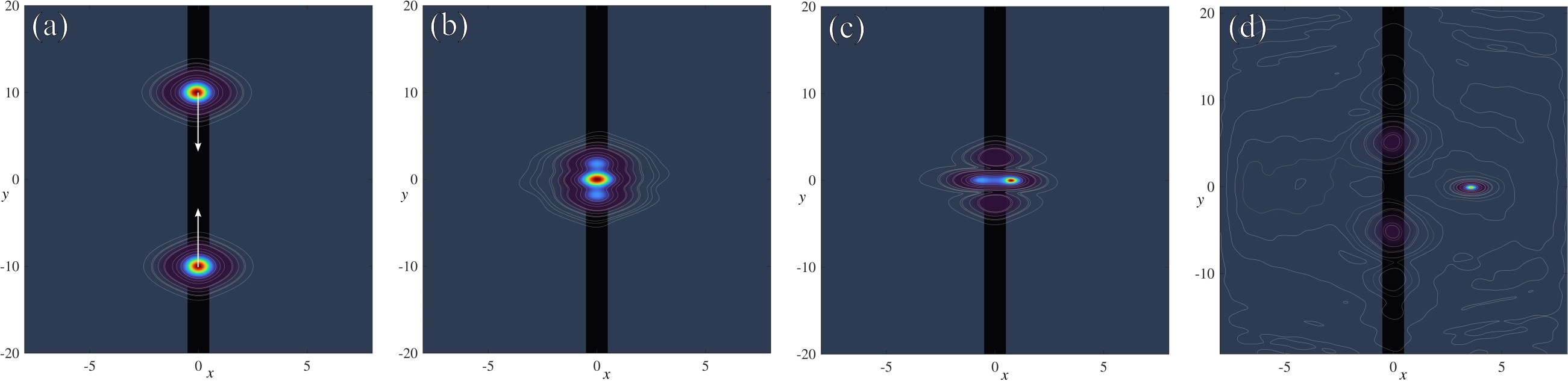}
    \caption{Breakup state from direct simulations at $N=8$, $v=0.8$ in Fig.~\ref{fig:7}(a). Panels (a)--(d) follow $|\Psi(\mathbf{r},z)|$ from the pre-collision stage through loss of confinement and decay. The HVA does not reproduce this process accurately.}
    \label{fig:10}
\end{figure*}

Increasing $N$ brings in the Breakup state. In the direct simulation of Fig.~\ref{fig:10}, strong overlap produces a transient two-peak structure along the confined $Ox$ direction, similar to the unstable state in Fig.~\ref{fig:5}. The field then rapidly loses transverse localization and escapes from the guiding channel. Most radiation is emitted along $Ox$, while the remaining field also spreads along the free $Oy$ direction.

This breakup mechanism lies outside the adopted HVA ansatz, which constrains the $Ox$ dependence to a Gaussian profile and places the principal dynamical freedom along $Oy$.

The stationary calculations in Fig.~\ref{fig:3}(a) show that stable states exist only below approximately $(N,\mu)<(16,1.48)$. Above this threshold, the soliton develops a two-peak structure and decays from the guiding channel during propagation, as illustrated in Fig.~\ref{fig:5}(b)--(c). During a collision, overlap of the two solitons can raise the combined power to the critical level, making the field unstable during the interaction and causing it to decay out of the channel. Thus, at large $N$ the Breakup region appears as a sharp ``wall'' in the $(N,v)$ map of Fig.~\ref{fig:7}(a).

Other normalized parameter sets $(\alpha,\gamma_{\mathrm{cl}},\sigma_{\mathrm{c}})$ below the boundary surface produce collision maps with the same qualitative structure as Fig.~\ref{fig:7}(a).

For the type-(II) region, the $(N,v)$ diagram is shown in Fig.~\ref{fig:7}(b). Across almost the entire range considered, the HVA agrees well with direct simulations of Eq.~\eqref{eq:5}. The four outcomes--Molecular, Oscillation, Splitting, and Breakup--and their boundaries are reproduced with good accuracy. A second parameter set above the boundary surface gives the same qualitative HVA map, supporting the generality of this picture.

\begin{figure}[!tb]
    \centering
    \includegraphics[width=\columnwidth]{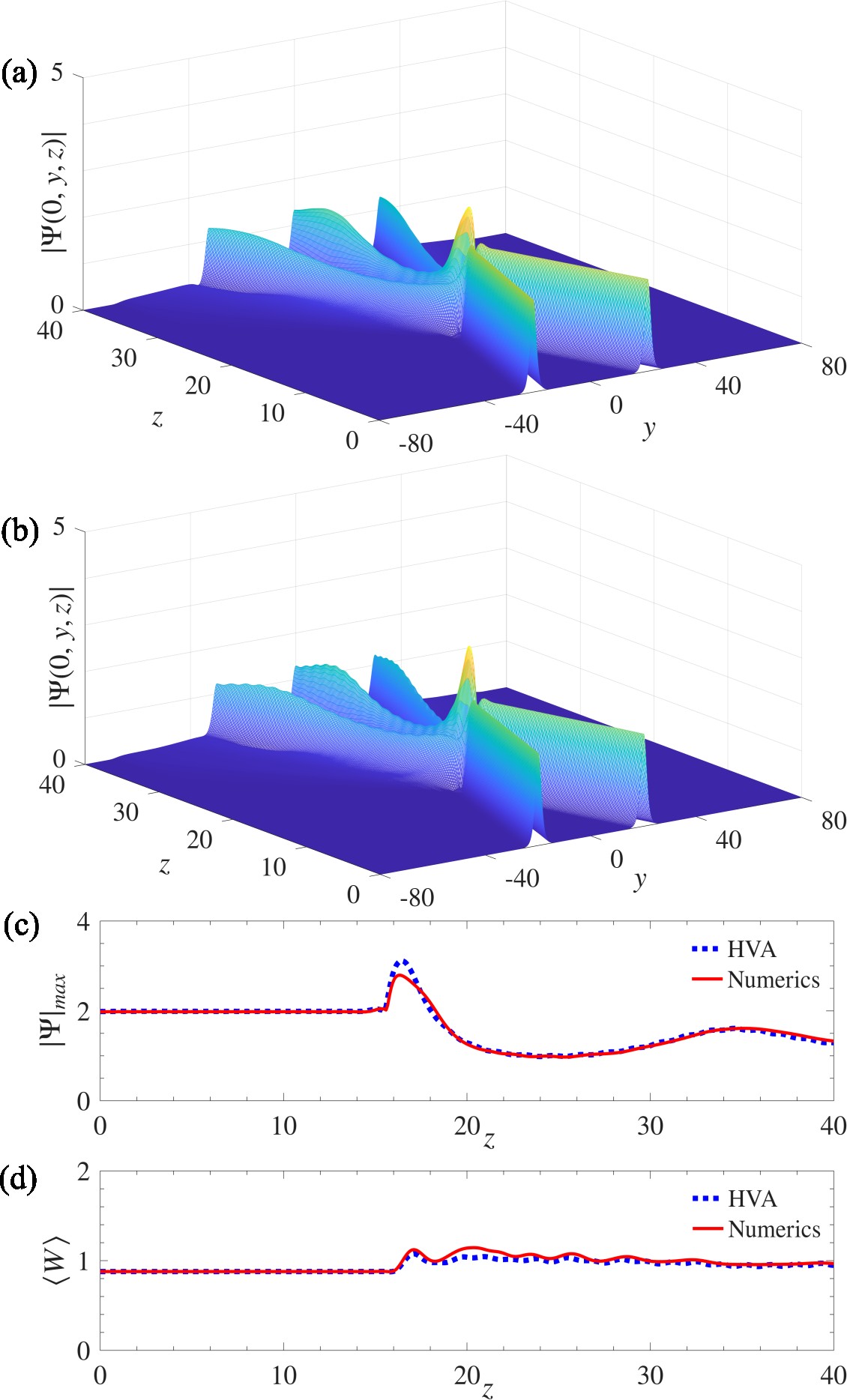}
    \caption{Splitting state in Fig.~\ref{fig:7}(a) with $N_0=9$ and $v=1.2$. Panels (a) and (b) show the $yz$ evolution of $|\Psi(\mathbf{r},z)|$ from direct simulations and the HVA ($n=1$), respectively; panels (c) and (d) compare $|\Psi(\mathbf{r},z)|_{\max}$ and $\langle W(z)\rangle$.}
    \label{fig:11}
\end{figure}

The HVA remains accurate for Molecular states as the power increases into the flat-top regime; compare the right column of Fig.~\ref{fig:8}. The high-power bound state, however, behaves differently from its low-power counterpart. After merger, $|\Psi(z)|_{\max}$ and $\langle W(z)\rangle$ undergo long-lived, nearly periodic oscillations rather than rapidly settling to constants. Radiation is weak and is emitted mainly during the initial collision and merger; afterwards, the bound state propagates with little further radiation.

Figure~\ref{fig:7}(b) contains two distinct Oscillation regions. In the larger one, for $N<17$, the collision is nearly elastic and resembles the low-power example in Fig.~\ref{fig:9}. For $N>17$, the stationary profiles evolve toward flat-top states and the collision becomes inelastic: the outgoing velocity changes, while $|\Psi|_{\max}$ and $\langle W(z)\rangle$ show persistent, nearly periodic oscillations, indicating strong excitation. Similar behavior occurs in the smaller high-power Oscillation region between Splitting and Molecular. The HVA captures both regimes and follows their numerical boundaries closely, as indicated by the dashed curves in Fig.~\ref{fig:7}(b).

The Splitting state appears for sufficiently deep confinement and is characterized by breakup of the initial structure into three or more localized components. These secondary packets either propagate or remain nearly stationary along the free $Oy$ direction and persist as breathers, with relatively little radiation. The process is therefore strongly inelastic. In the low-power example of Fig.~\ref{fig:11}, the collision produces three components: one remains near the center and two move in opposite directions. The HVA reproduces the direct simulation closely, apart from small additional oscillations superimposed on the breathing motion. At high power, collisions of flat-top solitons generate more secondary components, yet the HVA still captures the splitting process and its domain in parameter space reasonably well.

For $N>30$ and $v>3$, the system enters the Breakup region. Radiation is emitted continuously in all directions, more strongly along $Ox$ than along $Oy$, and the localized structures are eventually destroyed. Because the HVA retains its main dynamical freedom along $Oy$, it cannot reproduce the detailed two-dimensional radiation pattern. It nevertheless predicts breakup qualitatively, and the corresponding HVA boundary in Fig.~\ref{fig:7}(b) remains close to the numerical one.

\section{Conclusions}
We have studied spatial optical solitons and their interactions in a single-channel planar waveguide with spatially modulated refractive index and cubic--quintic nonlinearity, using the VA, HVA, and direct numerical simulations.

For stationary states, the Gaussian VA yields two types of $N(\mu)$ curves, where $N$ is soliton norm and $\mu$ is propagation constant, separated by a boundary surface in parameter space. The position of this surface is strongly influenced by the linear-potential depth $\alpha$, i.e., by the core--cladding index contrast. Numerical calculations show the same two qualitative types. Below the surface, VA and numerics agree at low power but diverge at high power, where the numerical state becomes two-peaked and unstable. Above the surface, both approaches retain the same qualitative $N(\mu)$ form. In this region, a super-Gaussian ansatz markedly improves the high-power description and accurately represents flat-top solitons.

The collisions produce four post-interaction states: Oscillation, Molecular, Splitting, and Breakup. The HVA reproduces most of the dynamical map and its boundaries. Above the stationary-state boundary surface, this agreement persists over almost the entire power range, including the flat-top regime. Below the surface, the HVA remains accurate for low-power Oscillation and Molecular states but fails for the high-power Breakup process, where the field develops a two-peak transverse structure and radiates strongly out of the channel.

These results emphasize that the reliability of the variational reduction is controlled by both the ansatz and the strength of transverse confinement. The boundary surface obtained from the Gaussian VA not only separates the two types of $N(\mu)$ curves but also remains meaningful for the full numerical model, providing a qualitative indicator of where the HVA can describe the collision dynamics reliably. The HVA therefore offers an efficient route to complex soliton interactions at substantially lower computational cost than direct simulations. Possible extensions include asymmetric structures, nonlocal nonlinearities, and spatiotemporal solitons.

\section{Acknowledgments}
This research is funded by the Vietnam National Foundation for Science and Technology
Development (NAFOSTED) under grant number 103.01-2021.152.

{
\footnotesize
\setlength{\bibsep}{0pt}
\bibliographystyle{elsarticle-num}
\bibliography{references}
}

\end{document}